\documentclass[nojss,shortnames]{jss}
\usepackage{amsfonts}
\usepackage{amsmath}
\usepackage{tikz}
\usetikzlibrary{shapes.geometric, arrows, positioning}
\tikzstyle{io} = [rectangle, text width=5em,text centered, draw=black]
\tikzstyle{ana} = [rectangle, text width=5em,text centered, draw=black]
\tikzstyle{sum} = [rectangle, text width=7em,text centered, draw=black]
\tikzstyle{vis} = [rectangle, text width=9.4em,text centered, draw=black]
\tikzstyle{title} = [rectangle, text width=5em, text centered]
\tikzstyle{arrow} = [thick,->,>=stealth]

\newcommand{\R}{\mathbb{R}}

\title{\pkg{RESI}: An \proglang{R} Package for Robust Effect Sizes}
\Plaintitle{RESI: An R Package for Robust Effect Sizes}
\author{Megan Jones\\Vanderbilt University\And
Kaidi Kang\\Vanderbilt University\And Simon Vandekar\\Vanderbilt University}
\Plainauthor{Megan Jones, Kaidi Kang, Simon Vandekar}
\Address{
Megan Jones\\
Vanderbilt University\\
Department of Biostatistics\\
2525 West End Ave., Suite 1136, Nashville\\
Tennessee 37203, United States of America\\
E-mail: \email{megan.n.taylor@vanderbilt.edu}\\
}

\Abstract{
Effect size indices are useful parameters that quantify the strength of association and are unaffected by sample size. There are many available effect size parameters and estimators, but it is difficult to compare effect sizes across studies as most are defined for a specific type of population parameter. We recently introduced a new Robust Effect Size Index (RESI) and confidence interval, which is advantageous because it is not model-specific. Here we present the RESI R package, which makes it easy to report the RESI and its confidence interval for many different model classes, with a consistent interpretation across parameters and model types. The package produces coefficient, ANOVA tables, and overall Wald tests for model inputs, appending the RESI estimate and confidence interval to each. The package also includes functions for visualization and conversions to and from other effect size measures. For illustration, we analyze and interpret three different model types.
}


\begin{document}
\maketitle

\section{Introduction}

Standardized effect sizes are unitless indices used to describe the magnitude of an association. Unlike $p$~values, which are often used to evaluate statistical significance, effect sizes do not depend on sample size \citep{betensky_p-value_2019}. A well known criticism of $p$~values and significance testing is that for large sample sizes, very small effects will be found as significant, even though these effects may be negligible in real-world application \citep{wasserstein_asas_2016}. In contrast, effect sizes communicate the strength of the effect rather than the existence of an effect of arbitrary size, which may be more meaningful in practice \citep{sullivan_using_2012}. Although increased sample size helps improve the precision of the estimate of an effect size, the effect size is a parameter that is not dependent on sample size \citep{kang_accurate_2023}. Journals and statistical guidelines are increasingly encouraging authors to report effect sizes and their CIs alongside or in place of $p$~values \citep{wasserstein_asas_2016, wilkinson_statistical_1999, american_psychological_association_publication_1994,
american_psychological_association_publication_2001, american_psychological_association_publication_2010,  althouse_recommendations_2021}. However, they are still not commonly reported \citep{fritz_effect_2012, amaral_current_2021} and when reported, they often do not include confidence intervals \citep{fritz_effect_2012}. 

There are four challenges to reporting effect sizes that limit their widespread use. First, there are many different effect size measures available \citep{cohen_statistical_1988, hedges_statistical_1985, rosenthal_parametric_1994, zhang_robust_1997, serdar_sample_2021}, but they are typically defined in the context of a specific population parameter, which makes comparing effects across a wide range of models difficult \citep{vandekar_robust_2020}. Second, many available effect size measures do not allow for nuisance parameters or covariates \citep{vandekar_robust_2020}. Third, many effect size measures do not have accurate confidence interval procedures, which precludes quantification of the uncertainty around the effect size estimate \citep{kang_accurate_2023}. Finally, many default model summary functions available in statistical software automatically output $p$~values, but few also report effect sizes.
The \pkg{RESI} \proglang{R} package was designed to address these challenges.

There are several \proglang{R} packages available for effect size calculation. For example, packages such as \pkg{MOTE}, \pkg{MBESS}, \pkg{effsize}, \pkg{esvis}, and \pkg{rcompanion} include functionality that allows the user to manually input data or the relevant test statistics for conversion to a desired effect size measure \citep{buchanan_mote_2019, kelley_mbess_2022, torchiano_effsize_2020, anderson_esvis_2020, mangiafico_rcompanion_2023}. The \pkg{effectsize} package implements many effect size measures and conversions between some of them \citep{ben-shachar_effectsize_2020}. \pkg{effectsize} allows users input statistics and models directly to compute the desired effect size. Although these tools are available, they do not address the general challenges to reporting and comparing effect sizes mentioned above. There is a need for an effect size index that can be broadly applied across model types. Additionally, user-friendly software tools that implement such a measure are needed to promote easy reporting of effect sizes.

The recently proposed Robust Effect Size Index (RESI) \citep{vandekar_robust_2020,kang_accurate_2023} addresses many of these challenges because it is broadly applicable across all common model types, it accommodates nuisance parameters, and there is an effective confidence interval procedure available \citep{kang_accurate_2023}. The RESI can be estimated from Chi-square, $F$, $Z$, and $t$ statistics. It is also possible to convert RESI estimates to and from other common effect size measures, such as Cohen's \textit{d}, Cohen's $f^2$, and $R^2$ \citep{vandekar_robust_2020}.
%Suggested meaningful ranges of the RESI are derived from Cohen's \textit{d} \citep{cohen_statistical_1988, vandekar_robust_2020}. 

The \pkg{RESI} \proglang{R} package builds on existing infrastructure for robust standard error estimation \citep{zeileis_object-oriented_2006} allowing easy estimation, reporting, and visualization \citep{jones_resi_2023}. Similarly to the \pkg{effectsize} package, \pkg{RESI} is designed to work on model inputs, so that effect size estimates can be easily obtained in tandem with common model summaries. These model-based functions also allow for a large amount of customization in the estimation and reporting process. Directly inputting test statistics and the relevant degrees of freedom and sample size is an option as well, helpful for model types that have not yet been implemented via dedicated methods in the package. The package also aims to work with other effect size measures, providing functions to convert to and from a few common effect size indices. Plotting functions are provided to allow for quick visualization of the effect size estimates present in models. With these tools, we hope to make obtaining the highly generalizable RESI simple and accessible, in order to increase ease of reporting effect sizes in research. In this paper, we outline the theory underlying the RESI, its estimators, and confidence interval procedure. We then discuss the \pkg{RESI} package, its structure, function arguments, and dependencies. Finally, we provide three in-depth examples of using the \pkg{RESI} package to perform analysis of effect sizes, from model creation to post-estimation visualization.

\section{Statistical methods}
\subsection{RESI definition}

The RESI is defined from the noncentrality parameter of a test statistic in the context of M-estimation, so it is broadly applicable across statistical models and parameters.
A full introduction to the RESI can be found in our previous work \citep{vandekar_robust_2020,kang_accurate_2023}. 
% Consider an estimating equation, 
% \begin{equation}
%     \Psi(\theta ; W) =   n^{-1}\sum_{i=1}^n \psi(\theta ; W)
% \end{equation}

% where $\Psi \in \mathbb{R}$ and $\psi$ is a known function. We obtain the M-estimator $\hat{\theta}$ as the maximizer of the estimating equation:

% \begin{equation}
%     \hat{\theta} = \arg\max\limits_{\theta^* \in \Theta} \Psi(\theta^* ; W)
% \end{equation}

% Under regularity conditions \citep{van_der_vaart_asymptotic_2000,vandekar_robust_2020}, the asymptotic robust covariance of $\sqrt{n} (\hat{\theta} - \theta)$ is an $m$ x $m$ matrix 

% \begin{equation}
%     \Sigma_{\theta} = \mathbf{J}(\theta)^{-1}\mathbf{K}(\theta)\mathbf{J}(\theta)^{-1}
% \end{equation}

% where

% \begin{equation*}
%     \mathbf{J}_{jk}(\theta) = -\lim\limits_{n \to \infty} E \frac{\partial^2 \Psi (\theta^*;W)}{\partial \theta_j^*\partial\theta_k^*} \biggr\rvert_{\theta}
% \end{equation*}

% \begin{equation*}
%     \mathbf{K}_{jk}(\theta) = \lim\limits_{n \to \infty} E \frac{\partial \Psi (\theta^*;W)}{\partial \theta_j^*} \frac{\partial \Psi(\theta^*; W)}{\partial \theta_k^*} \biggr\rvert_{\theta}
% \end{equation*}
Briefly, consider a dataset of independent observations $W = \{W_1, \ldots, W_n\}$ with probability distribution $\mathcal{P}$ and let $\theta = (\alpha, \beta) \in \mathbb{R}^m$ be a vector of parameters with $\alpha \in \mathbb{R}^{m_0}$ nuisance parameters and $\beta \in \mathbb{R}^{m_1}$ the target parameters of interest.
The RESI is constructed using the test statistic for the null hypothesis $H_0: \beta_0 \in \R^{m_1}$, where $\beta_0$ is a reference value, usually zero \citep{vandekar_improving_2021}.
Assuming known variance, the usual Wald-style test statistic, centered at the reference value, $T^2 = n(\hat{\beta} -\beta_0)^T\Sigma_\beta^{-1}(\hat{\theta})(\hat{\beta} -\beta_0)$ follows a  Chi-square distribution with $m_1$ degrees of freedom and noncentrality parameter $n(\beta - \beta_0)^T\Sigma_\beta^{-1}(\beta - \beta_0)$. The RESI, $S_\beta$, is the square root of the component of the noncentrality parameter that does not depend on the sample size

\begin{equation*}\label{eq:sbeta}
    S_\beta = \sqrt{(\beta - \beta_0)^\top\Sigma_\beta^{-1}(\beta - \beta_0)}.
\end{equation*}

\subsection{RESI estimators} \label{sec:resi estimators}

The RESI is very general, because its estimator can be computed for Chi-square, $F$, $Z$, and $t$ statistics as well \citep{vandekar_robust_2020,kang_accurate_2023}. In this section, we review these estimators and introduce new estimators for a modified RESI using $Z$ and $t$ statistics, which have the advantage that the proposed modification shows the direction of the effect for univariate parameters. We also describe the use of robust covariance in the estimation of the test statistics.

The original estimator for $S_\beta$ was developed using an estimator for noncentrality parameters of Chi-square statistics \citep{vandekar_robust_2020}

\begin{equation}\label{eq:chisq2S}
    \hat{S}_\beta = \biggr\{\max\biggr[0, \frac{T^2 - m_1}{n}\biggr]\biggr\}^{\frac{1}{2}}.
\end{equation}
Because $S_\beta$ is nonnegative, the max operator ensures that the estimator is also nonnegative in finite samples.

Under normality, the finite sample distribution of the asymptotic Chi-square statistic divided by its degrees of freedom is an F distribution \citep{mantel_chi-square_1963}. When this is true, a better small sample estimator can be computed using method of moments with the F distribution

\begin{equation}\label{eq:F2S}
    \hat{S}_\beta = \biggr\{\max\biggr[0, \frac{F\times (n-m-2) - m_1\times (n-m)}{n \times(n-m)}\biggr]\biggr\}^{\frac{1}{2}}.
\end{equation}

The RESI is called robust because its estimator is consistent under model misspecification when estimated with a robust test statistic, \citep{vandekar_robust_2020,mackinnon_heteroskedasticity-consistent_1985}.
%When the covariance is unknown in linear regression, a common choice for estimation is a robust (sandwich) covariance estimator of the form $(X^TX)^{-1}X^T\Omega X(X^TX)^{-1}$, where $\Omega$ is a known function that differs based on the given method \citep{white_heteroskedasticity-consistent_1980,mackinnon_heteroskedasticity-consistent_1985,long_using_2000}.
which uses a heteroskedastic consistent sandwich estimator for $\Sigma_\beta$ \citep{white_heteroskedasticity-consistent_1980,mackinnon_heteroskedasticity-consistent_1985,long_using_2000}.
The RESI estimator is a consistent estimator of the true effect size, even if there is unknown heteroskedasticity between the measurements.
When the mean model is misspecified, the RESI is a consistent estimator of the best approximation of the true model within the class of models considered \citep{boos_essential_2013}.

With equations~\eqref{eq:chisq2S} and \eqref{eq:F2S}, we can compute RESI estimates for Chi-square and $F$ statistics, which are easily obtained from many statistical models. However, these estimates have the feature of being nonnegative, so they do not describe the direction of an effect.
While this makes them generally applicable across univariate (can be negative and positive) and multivariate (can only be positive) parameters, for univariate parameters it is also useful to be able to obtain a signed effect size estimate, showing the directionality of the effect. With this in mind, we introduce RESI estimators for $Z$ and $t$ statistics. We use two approaches to develop these estimators, leading to two estimators with different theoretical properties and advantages.

The first approach is the same as the development of the RESI estimators for  Chi-square and $F$ statistic. We use the method of moments for the $Z$ or $t$ statistics to find estimators for $S_\beta$. 
Consider a $Z$ statistic, whose expected value is $\E Z= \sqrt{n}\mathrm{sgn}(\beta) S_\beta$, where $\mathrm{sgn}$ is the sign function, which leads to the signed RESI estimator

\begin{equation}\label{eq:z2S}
    \hat{S}_\beta = \frac{Z}{\sqrt{n}}
\end{equation}

For a $t$ statistic with degrees of freedom $n-m$ and noncentrality parameter $\sqrt{n}S_\beta$, when $n - m > 1$, the expected value is $\E t = \sqrt{\frac{n(n-m)}{2}} \frac{\Gamma((n-m-1)/2)}{\Gamma((n-m)/2)} S_\beta  $. This gives the RESI estimator

\begin{equation}\label{eq:t2S}
    \hat{S}_\beta = \frac{t\sqrt{2} \Gamma((n-m)/2)}{\sqrt{n(n-m)} \Gamma((n-m-1)/2)}
\end{equation}
The advantage of estimators \eqref{eq:z2S} and \eqref{eq:t2S} is that both are unbiased for $S$.

The second approach leverages the relationship between $Z$ and Chi-square statistics and $t$ and $F$ statistics. Squaring a $Z$ or $t$ statistic gives a Chi-square or $F$ statistic, respectively.
We then use equations~\eqref{eq:chisq2S} and \eqref{eq:F2S} as RESI estimators for $Z$ and $t$ statistics by multiplying them with the sign of the test statistic. For example,
\begin{equation}\label{eq:z2S_alt}
    \hat{S}_\beta = \mathrm{sgn}(Z)\times \biggr\{\max\biggr[0, \frac{Z^2 - 1}{n}\biggr]\biggr\}^{\frac{1}{2}},
\end{equation}
and similarly for the $t$ estimator.
These estimators are biased, but consistent and have smaller mean squared error than estimators~\eqref{eq:z2S} and \eqref{eq:t2S}. These estimators are advantageous because their estimates are equal in absolute value to the unsigned RESI estimates, whereas the estimators \eqref{eq:z2S} and \eqref{eq:t2S} are not.

\subsection{Bootstrapping procedure for confidence intervals}
\label{sec:bootstrapping}

In recent work, we showed that Chi-square and $F$ confidence intervals are not accurate for computing effect size confidence intervals in general.
In particular, when the test statistic is estimated using a robust covariance estimator, using a Chi-square or $F$ distribution for the RESI estimate underestimates the variance and will therefore produce confidence intervals that do not exhibit the nominal coverage level \citep{kang_accurate_2023}.
As an alternative, we proposed a nonparametric bootstrap for the RESI confidence interval \citep{kang_accurate_2023}. Because the nonparametric bootstrap confidence interval most consistently produces confidence intervals with nominal coverage, that is the default procedure in the RESI package. For linear models and nonlinear least squares models, a Bayesian bootstrap is also implemented as an option \citep{rubin_bayesian_1981}.

\subsection{Meaningful RESI ranges}
\label{sec:ranges}

When interpreting the RESI estimates, it is useful to have an idea of what constitutes a "large" or "small" effect. While a meaningful effect size ultimately depends on the scientific context, ranges can be posited based on published effect size ranges \citep{cohen_statistical_1988}. When the RESI was introduced, meaningful ranges were derived based on a conversion from Cohen's $d$ assuming equal sample proportions \citep{vandekar_robust_2020}.
The suggested range for no effect to small effect is a RESI of [0, 0.1]. The suggested range for a small effect to medium effect is a RESI of (0.1, 0.25]. A medium to large effect is suggested as a RESI of (0.25, 0.4], and large effects are greater than 0.4.

\section{The RESI package}

\pkg{RESI} is available to the public via \href{https://cran.r-project.org/package=RESI}{The Comprehensive R Archive Network (CRAN)}. To download, one can use the following code:

\begin{CodeChunk}
\begin{CodeInput}
R> install.packages("RESI")
\end{CodeInput}
\end{CodeChunk}

The development version is available on \href{https://github.com/statimagcoll/RESI}{GitHub}. This can be downloaded using the \pkg{devtools} package with the following command \citep{wickham_devtools_2022}:

\begin{CodeChunk}
\begin{CodeInput}
R> devtools::install_github("statimagcoll/RESI")
\end{CodeInput}
\end{CodeChunk}

\subsection{Operation}

\begin{figure}
    \centering
    \begin{tikzpicture}[node distance=2cm]
    \node (title1) [title] {\textbf{\underline{Inputs}}};
    \node (title2) [title, right=of title1, xshift=-0.8cm] {\textbf{\underline{Analysis/}} \\ \textbf{\underline{Computation}}};
    \node (title3) [title, right=of title2] {\textbf{\underline{Summaries}}};
    \node (title4) [title, right=of title3, xshift=0.58cm] {\textbf{\underline{Visualization}}};
    \node (input1) [io, below=of title1, yshift=1.5cm] {Fitted model object};
    \node (input2) [io, below=of input1, yshift=1cm] {$Z$, $t$, Chi-square, or $F$ statistic};
    \node (input3) [io, below=of input2, yshift=1cm] {Other effect size measure ($R^2$, $f^2$, $d$)};
    \node (input4) [io, below=of input3, yshift=1cm] {RESI (S) estimate};
    \node (analysis1) [ana, right=of input1, xshift=-0.5cm] {\code{resi} \\ \code{resi\textunderscore pe}};
    \node (analysis2) [ana, right=of input2, xshift=-0.5cm] {\code{z2S} \\ \code{z2S\textunderscore alt} \\ \code{t2S} \\ \code{t2S\textunderscore alt} \\ \code{chisq2S} \\ \code{f2S}};
    \node (analysis3) [ana, right=of input3, xshift=-0.5cm] {\code{Rsq2S} \\ \code{fsq2S} \\ \code{d2S}};
    \node (analysis4) [ana, right=of input4, xshift=-0.5cm] {\code{S2Rsq} \\ \code{S2fsq} \\ \code{S2d}};
    \node (summary1) [sum, right=of analysis1, xshift=-0.5cm] {\code{summary.resi} \\ \code{Anova.resi} \\ \code{anova.resi}};
    \node (visual1) [vis, right=of summary1, xshift=-0.5cm] {\code{plot.resi} \\ \code{plot.summary_resi} \\ \code{plot.anova\textunderscore resi} \\ \code{print.resi} \\ \code{print.summary_resi}};
    \node[inner sep=0pt] (logo) at (9.2,-6)
    {\includegraphics[width=.32\textwidth]{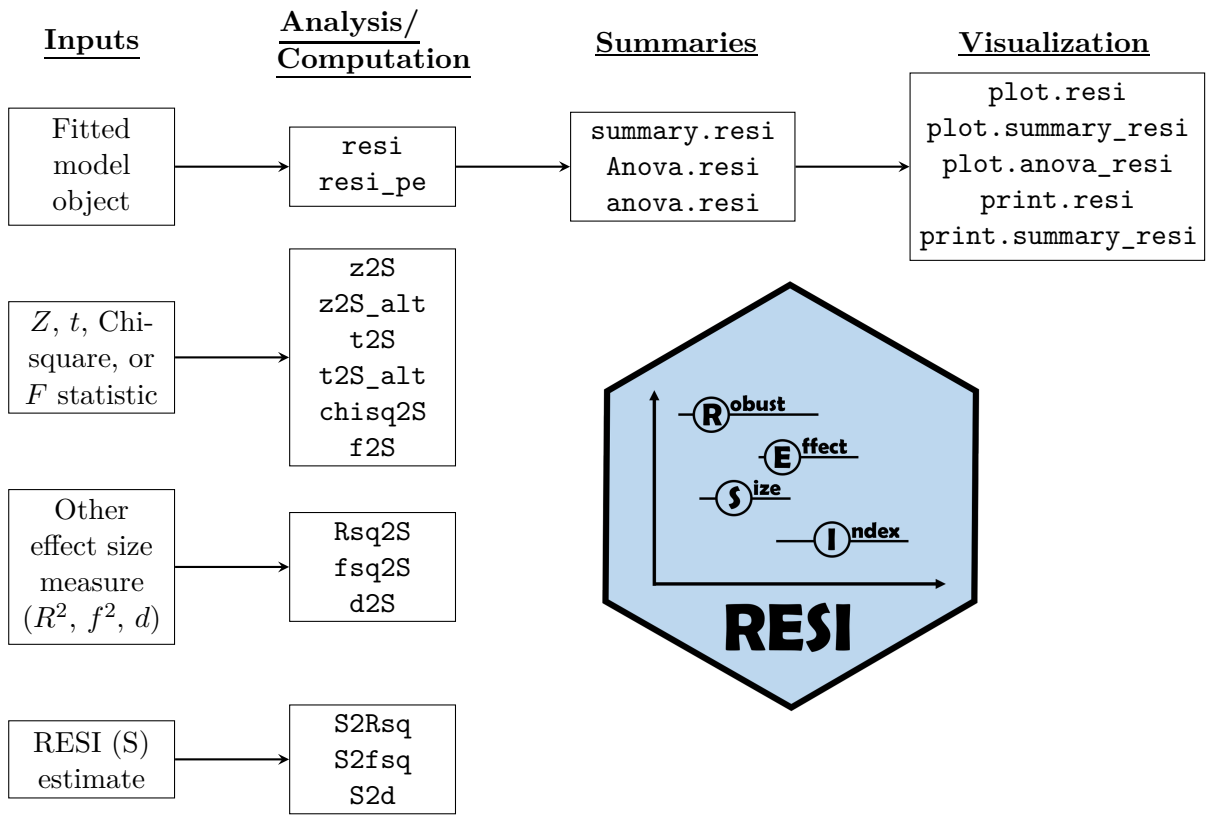}};
    \draw [arrow] (input1) -- (analysis1);
    \draw [arrow] (analysis1) -- (summary1);
    \draw [arrow] (summary1) -- (visual1);
    \draw [arrow] (input2) -- (analysis2);
    \draw [arrow] (input3) -- (analysis3);
    \draw [arrow] (input4) -- (analysis4);
    \end{tikzpicture}
    \caption{\pkg{RESI} Package Structure and Logo. Inputs to package functions can be models of supported types, test statistics with relevant degrees of freedom and sample size, or effect size measures. The analysis functions compute RESI estimates with or without confidence intervals, or convert to and from other effect size indices. Summary functions provide relevant information extracted from a `\code{resi}' object. Post-estimation visualization functions include plotting and printing.}
    \label{fig:pkg structure}
\end{figure}

Users should have \proglang{R} version 2.10 or higher to use \pkg{RESI} \citep{r_core_team_r_2022}.
The \pkg{RESI} package is designed to easily add RESI estimates and confidence intervals to common model outputs, such as coefficient summaries and ANOVA tables.
% The confidence intervals are obtained using bootstrapping \citep[see][for details]{kang_accurate_2023}. I guess this would go above in Section 2.
The functions in the package are split into three categories: model-based functions, conversion functions, and additional methods to other functions (Figure~\ref{fig:pkg structure}).
There are also two datasets provided.

\subsection{Model-based functions}

The main model-based RESI functions of the \pkg{RESI} package are \code{resi\textunderscore pe()}, to obtain point estimates, and \code{resi()} for point estimates with confidence or credible intervals. \code{resi\textunderscore pe()} uses standard summary and ANOVA outputs to compute the RESI point estimate. \code{resi()} uses \code{resi\textunderscore pe()} and performs bootstrapping to produce confidence intervals for the RESI. These functions take supported fitted models as input and return a list that contains three main components: a coefficients summary table with a row for each non-reference level of each variable, an ANOVA table containing a row for each variable, and an overall RESI estimate.
Details regarding functions used for table construction for the supported model types are given in Table~\ref{tab:model types}.

While the user can simply run \code{resi()} on a supported model type and obtain a full output, there are several arguments that can be used to tailor the process. Details for all function arguments are available in the documentation, but we briefly cover important arguments here. \code{resi()} and \code{resi\textunderscore pe()} both contain the following arguments. The \code{model.full} argument is the model to perform RESI estimation on. The \code{model.reduced} argument, \code{NULL} by default, specifies a reduced model which is used to compute an effect size estimate in comparison to the full model for a specific subset of variables that the user wishes to compare (See \hyperref[illustration 3]{RESI on survival model}). If left as \code{NULL}, \code{resi\textunderscore pe()} will compute a reduced model of the same type as the full model, but including only the intercept term. \code{data} is a blank argument referring to the data used to generate the model. If left blank, \code{resi()} pulls the data from the model. For some model types (`\code{survreg}', `\code{coxph}', `\code{nls}'), the data is required as an input because these models objects do not store the original data frame used to fit the model. Additionally, when using some formula functions such as splines or factoring, the data needs to be input so that the spline arguments can be recomputed as they were in the original data.

The \code{vcovfunc} argument can be used to specify a different variance-covariance function and is important because it affects whether the effect size is robust to model misspecification (see \hyperref[sec:resi estimators]{RESI Estimators}). By default, RESI will use a robust covariance estimator. Additional arguments to the given \code{vcovfunc} function can be specified in list form with the \code{vcov.args} argument. Similarly, additional arguments to the \code{Anova()} function (from \pkg{car} \citep{fox_r_2019}) can be specified with the \code{Anova.args} argument. The \code{unbiased} argument is logical (default \code{TRUE}) and corresponds to a choice of conversion formulas for the $Z$ and $t$ statistics (see \hyperref[sec:resi estimators]{RESI Estimators} for details).

\code{resi()} contains additional arguments related to the bootstrap procedure. The confidence level (default 0.05) for the confidence or credible intervals can be specified with \code{alpha}. Multiple confidence levels can be specified using a numeric vector. For `\code{lm}' and `\code{nls}' models, there is a \code{boot.method} argument that can be specified as nonparametric (default) or Bayesian (see \hyperref[sec:bootstrapping]{Bootstrapping}). Finally, the \code{store.boot} argument (default \code{FALSE}) determines whether to store the full table of bootstrapped estimates, which can be useful if the user wants to be able to obtain confidence intervals with different confidence levels without rerunning the bootstrap procedure.

The output of \code{resi()} is a list of class `\code{resi}' that contains the three main tables (coefficients, ANOVA, and overall) with confidence intervals and several other elements to track how the functions were called. \code{resi\textunderscore pe()} produces a list with these tables (without confidence intervals) and other elements about the model. 

The \code{overall} element of the output is a table reporting a Wald test comparing the full model to the reduced model. The test statistic is typically converted from a Chi-square statistic to a RESI estimate internally using the \code{chisq2S()} function, which takes the number of observations from the data and degrees of freedom from the Wald test. In the case of a linear model, the RESI estimate is computed using the \code{f2S()} function. 

The \code{coefficients} table is available for every model type supported by the package. The \code{coefficients} argument (default = \code{TRUE}) in \code{resi()}/\code{resi\textunderscore pe()} determines whether to compute this table. This provides a RESI estimate for each model coefficient and appends it to a table resulting from one of the ``Coefficients'' functions in Table~\ref{tab:model types}. The $Z$ or $t$ statistic from this function is converted to the signed RESI via either \code{z2S()}/\code{t2S()} for the unbiased version, or \code{z2S\textunderscore alt()}/\code{t2S\textunderscore alt()}, for the alternate version.

The \code{anova} table is computed via \code{Anova()} from \pkg{car} \citep{fox_r_2019} where available (for `\code{geeglm}' models, \code{anova} is used). The \code{anova} argument (default = \code{TRUE}) in \code{resi()}/\code{resi\textunderscore pe()} determines whether to compute this table. For `\code{lm}' models, an $F$-test is used. For the others, a Wald test is specified assuming Chi-square statistics. Other options can be passed to \code{Anova()} function via \code{Anova.args}. Note that the \code{test.statistic} argument is fixed in the \code{resi\textunderscore pe()} function, so supplying a different value for this argument will result in an error. Additionally, if the user wishes to use a different \code{vcov.} argument in \code{Anova()} function, this should be done by providing the function to the \code{vcovfunc} argument in \code{resi()} (see \hyperref[illustration 1]{RESI on linear model}). Specifying this argument in \code{Anova.args} will result in an error. The resulting Chi-square or $F$ statistics are converted to RESI estimates using \code{chisq2S()} or \code{f2S()}. 

\code{RESI} for longitudinal models is still in development. Currently, the package provides point estimate and confidence interval methods for `\code{gee}' (from \pkg{gee} package \citep{carey_gee_2022}) and `\code{geeglm}' (from \pkg{geepack} \citep{halekoh_r_2006}) models. For these models, both a longitudinal RESI and a per-measurement cross-sectional RESI estimate are computed for each factor in the \code{coefficients} table (for `\code{gee}' and `\code{geeglm}') and for each variable in the \code{anova} table (for `\code{geeglm}'). The longitudinal RESI is the estimated effect conditional on the sampling design, whereas the cross-sectional estimator is the effect if the data were collected cross-sectionally. This allows investigators to quantify the benefit conferred by considering a longitudinal mode. For linear mixed effects models fit via \code{lme()} from \pkg{nlme} \citep{pinheiro_nlme_2021} and \code{lmerMod()} from \pkg{lme4} \citep{bates_fitting_2015}, longitudinal RESI point estimation is available in both a \code{coefficients} and \code{anova} table. The confidence interval procedure is still being evaluated for these models, so running \code{resi()} on a model of this type will provide point estimates only with a corresponding message. 

\begin{table}[t!]
\centering
\begin{tabular}{llllllp{7.4cm}}
\hline
Model           & Package & Covariance & Coefficients & Anova & Overall \\ \hline
`\code{lm}' & \pkg{stats} & \code{sandwich::vcovHC} & \code{coeftest} & \code{car::Anova} & \code{waldtest} \\
`\code{glm}' & \pkg{stats} & \code{sandwich::vcovHC} & \code{coeftest} & \code{car::Anova} & \code{waldtest} \\
`\code{nls}' & \pkg{stats} & \code{regtools::nlshc} & \code{coeftest} & N/A & \code{wald.test}\\
`\code{survreg}' & \pkg{survival} & \code{vcov} & \code{coeftest} & \code{car::Anova} & \code{waldtest} \\
`\code{coxph}' & \pkg{survival} & \code{vcov} & \code{coeftest} & \code{car::Anova} & \code{wald.test} \\
`\code{hurdle}' & \pkg{pscl} & \code{sandwich::sandwich} & \code{coeftest} & N/A & \code{waldtest} \\
`\code{zeroinfl}' & \pkg{pscl} & \code{sandwich::sandwich} & \code{coeftest} & N/A & \code{waldtest}\\
`\code{gee}' & \pkg{gee} & internal* & \code{summary} & N/A & N/A \\
`\code{geeglm}' & \pkg{geepack} & \code{vcov} & \code{coeftest} & \code{anova} & N/A \\
`\code{lme}' & \pkg{nlme} & \code{clubSandwich::vcovCR} & \code{summary} & \code{car::Anova} & N/A \\
`\code{lmerMod}' & \pkg{lme4} & \code{clubSandwich::vcovCR} & \code{summary} & \code{car::Anova} & N/A\\
\hline
\end{tabular}
\caption{\label{tab:model types} Supported model types and related functions. \code{coeftest} and \code{waldtest} are from \pkg{lmtest} \citep{zeileis_diagnostic_2002}. \code{wald.test} is from \pkg{aod} \citep{lesnoff_aod_2012}. *The robust covariance is computed within the \code{resi\textunderscore pe} function.}
\end{table}

\nocite{jackman_pscl_2020, pustejovsky_clubsandwich_2022, matloff_regtools_2022}

\subsection{Other package elements}

The package includes \code{print()}, \code{plot()}, \code{summary()}, \code{anova()}, and \code{car::Anova()} methods for `\code{resi}' objects. The \code{summary()} and \code{anova()}/\code{Anova()} methods are intended to isolate the corresponding elements of the `\code{resi}' object and allow the user to specify a different confidence level without having to rerun the bootstrapping process, if the \code{store.boot} option was set to \code{TRUE} when running \code{resi()}
%, the user can provide a vector of new $\alpha$ values using the \code{alpha} argument in these functions.
Running \code{summary()} on a `\code{resi}' object returns the \code{coefficients} table as an object of class `\code{summary\textunderscore resi}', with its own \code{plot()} and \code{print()} methods. Running \code{anova()} or \code{car::Anova()} on a `\code{resi}' object returns the \code{anova} table as an object of class `\code{anova\textunderscore resi}' and inherited classes from \code{anova()}/\code{car::Anova()}. There is also a \code{plot()} method for `\code{anova\textunderscore resi}'.

The package also contains a few conversion functions from RESI to and from other common effect size measures. These are Cohen's \textit{d}, Cohen's $f^2$, and $R^2$. Formulas for these conversions are found in \citep{vandekar_robust_2020}.

Lastly, the \pkg{RESI} package contains two datasets. The \code{insurance} dataset is adapted from the open-source repository Kaggle (\href{https://www.kaggle.com/datasets/teertha/ushealthinsurancedataset/discussion}{US Health Insurance Dataset}) and the \code{depression} dataset is adapted from a data analysis textbook \citep{agresti_categorical_2002}. Full details on the datasets are provided in the \pkg{RESI} package documentation.

\subsection{Important dependencies} \label{sec:important dependencies}

The \pkg{RESI} package currently has dedicated methods for 11 model types (Table~\ref{tab:model types}). The software function used to compute the covariance matrix varies by model type. It is possible to pass additional arguments to these covariance functions in \code{resi} by using the \code{vcov.args} argument. Any other valid covariance function can be specified as well. Although robust covariance estimators are used as the default for most model types, the survival models (`\code{survreg}' and `\code{coxph}') have the option for a robust covariance estimate in model setup and, when using the standard \code{vcov} from \pkg{stats}, they compute robust covariance matrices if the argument \code{robust=TRUE}. For `\code{geeglm}' models, the robust covariance is taken from the model directly \citep{zeileis_object-oriented_2006}.

Several other common analysis functions are used to obtain test statistics for RESI computation for the coefficients, ANOVA, and overall table. The functions used for different model types are found in Table~\ref{tab:model types}.

\section{Illustrations}

To demonstrate the flexibility of the \pkg{RESI} package, we analyze a few example datasets for several different model types using different covariance estimator functions and bootstrapping options. 

\subsection{RESI on linear model} \label{illustration 1}

We first look at a linear model fit using \code{lm()}. After installing the package from CRAN or GitHub, we load the \pkg{RESI} library.

\begin{CodeChunk}
\begin{CodeInput}
R> library("RESI")
\end{CodeInput}
\end{CodeChunk}

We will use the \code{insurance} dataset in the package to fit our model. The dataset contains information on insurance charges, age, sex, BMI, number of children, smoking status, and geographical region for 1338 individuals in the United States. We fit a linear regression of charges against region, age, BMI, and sex, with an interaction term on region and age and return the standard coefficients table using the summary function. 

\begin{CodeChunk}
\begin{CodeInput}
R> mod_lm <- lm(charges ~ region * age + sex + bmi, data = insurance)
R> summary(mod_lm)
\end{CodeInput}
\begin{CodeOutput}
Call:
lm(formula = charges ~ region * age + sex + bmi, data = insurance)

Residuals:
   Min     1Q Median     3Q    Max 
-14871  -7062  -4885   6235  46347 

Coefficients:
                    Estimate Std. Error t value Pr(>|t|)    
(Intercept)         -5359.44    2369.09  -2.262   0.0238 *  
regionnorthwest     -2339.44    2647.85  -0.884   0.3771    
regionsoutheast     -3230.85    2583.12  -1.251   0.2112    
regionsouthwest      -232.48    2662.84  -0.087   0.9304    
age                   220.33      45.08   4.888 1.14e-06 ***
sexmale              1328.02     622.07   2.135   0.0330 *  
bmi                   323.77      53.72   6.027 2.17e-09 ***
regionnorthwest:age    34.90      63.55   0.549   0.5829    
regionsoutheast:age    83.64      61.65   1.357   0.1751    
regionsouthwest:age   -33.63      63.74  -0.528   0.5979    
---
Signif. codes:  0 '***' 0.001 '**' 0.01 '*' 0.05 '.' 0.1 ' ' 1

Residual standard error: 11360 on 1328 degrees of freedom
Multiple R-squared:  0.126,	Adjusted R-squared:  0.1201 
F-statistic: 21.27 on 9 and 1328 DF,  p-value: < 2.2e-16
\end{CodeOutput}
\end{CodeChunk}

The $p$~values in the standard model summary indicate that age, sex, and BMI are significantly associated with insurance charges. However, just by looking at the $p$~values, it is hard to discern the strength of the the association. We would like to be able to see, in addition to significance, a measure of the effect size. To accomplish this, we can run \code{resi()} on the model object. We run it using all the default options first. This will use the \code{vcovHC()} function from the \pkg{sandwich} package (with default arguments) to compute robust standard error estimates \citep{zeileis_various_2020}. Since we are using the \code{resi()} function rather than the \code{resi\textunderscore pe()} function, we will obtain bootstrapped confidence intervals in addition to RESI point estimates. We set the seed to ensure the results are reproducible. This function can take several seconds to run. Printing the full `\code{resi}' object will print several tables and notes, so to begin we just print the summary.

\begin{CodeChunk}
\begin{CodeInput}
R> set.seed(0826)
R> resi_obj_lm <- resi(mod_lm)
R> summary(resi_obj_lm)
\end{CodeInput}
\begin{CodeOutput}
Analysis of effect sizes based on RESI:
Confidence level =  0.05
Call:  lm(formula = charges ~ region * age + sex + bmi, data = insurance)

Coefficient Table 
                      Estimate Std. Error t value Pr(>|t|)    RESI    2.5%  
(Intercept)         -5359.4352  2175.9439 -2.4630   0.0139 -0.0673 -0.1199 
regionnorthwest     -2339.4433  2395.1507 -0.9767   0.3289 -0.0267 -0.0800 
regionsoutheast     -3230.8512  2643.1099 -1.2224   0.2218 -0.0334 -0.0842 
regionsouthwest      -232.4839  2574.2823 -0.0903   0.9281 -0.0025 -0.0566 
age                   220.3325    40.2091  5.4797   0.0000  0.1497  0.0953  
sexmale              1328.0215   621.7421  2.1360   0.0329  0.0584  0.0079  
bmi                   323.7725    58.0849  5.5741   0.0000  0.1523  0.1049  
regionnorthwest:age    34.9040    57.2364  0.6098   0.5421  0.0167 -0.0360  
regionsoutheast:age    83.6359    63.3258  1.3207   0.1868  0.0361 -0.0176 
regionsouthwest:age   -33.6290    61.4065 -0.5476   0.5840 -0.0150 -0.0686 
                         97.5%
(Intercept)            -0.0111
regionnorthwest         0.0245
regionsoutheast         0.0170
regionsouthwest         0.0515
age                     0.2137
sexmale                 0.1146
bmi                     0.1983
regionnorthwest:age     0.0699
regionsoutheast:age     0.0884
regionsouthwest:age     0.0386
\end{CodeOutput}
\end{CodeChunk}

This output shows the \code{coefficients} element of the `\code{resi}' object, as well as the model call and the confidence level ($\alpha$, by default = 0.05). The coefficient table looks very similar to the standard model summary output. The estimates will remain unchanged, but the standard errors differ because \code{summary()} uses the model-based (naive) standard error, whereas \code{resi()} defaults to use a robust estimate. These standard error estimates will remain valid under heteroskedasticity. Accordingly, the $t$-values and $p$~values are different, but our qualitative conclusions about statistical significance are unchanged in this example. The three rightmost columns are new and represent the RESI estimates and $(1-\alpha)\%$ confidence intervals. Note that a RESI estimate further from 0 indicates a larger effect. The sign of the RESI estimate indicates the direction of the effect. From the table we can see that BMI is estimated to have a small to moderate effect (0.1523 (CI: 0.1049, 0.1983)) based on the ranges given in \hyperref[sec:ranges]{Meaningful RESI Ranges}. Sex is estimated to have a small effect (0.0584 (CI: 0.0079, 0.1146)). The effect size estimates are conditional on the other terms in the model. Because our model includes an interaction on age and region, the RESI estimate for ``age'' in the coefficient table is interpreted as the estimated effect size of age for those in the northeast (the reference region). This is estimated to be 0.1497 (CI: 0.0953, 0.2137), a small to moderate effect. For these results, if the $p$~value is less than 0.05, then the CI for the RESI does not contain 0. This will not always be the case because the RESI CI is estimated for the distribution of the effect size estimator under the alternative.

Because region is a factor variable, the test for region and its interaction with age corresponds to multiple parameters in the model. To obtain an effect size estimate for multiple parameters that correspond to a single variable, we can report the ANOVA table. We can obtain this with either the standard \code{anova()} function or the \code{car::Anova()} function on the `\code{resi}' object.

\begin{CodeChunk}
\begin{CodeInput}
R> anova(resi_obj_lm)
\end{CodeInput}
\begin{CodeOutput}
Analysis of Deviance Table (Type II tests)

Response: charges
           Df        F  Pr(>F)     RESI    2.5%   97.5%
region      3   1.5959 0.18856 0.036480 0.00000 0.11687
age         1 117.7046 0.00000 0.295111 0.23967 0.36204
sex         1   4.5624 0.03286 0.051549 0.00000 0.11128
bmi         1  31.0708 0.00000 0.149798 0.10130 0.19633
region:age  3   1.1167 0.34115 0.016056 0.00000 0.10525
\end{CodeOutput}
\end{CodeChunk}

By default, \code{resi()} uses a Type II sum of squares, but this can be changed in the arguments \citep{papachristodoulou_tutorial_2005}. This output is the same as running \code{car::Anova()} on the model using \code{sandwich::vcovHC} as the \code{.vcov} argument, but with the three rightmost columns added for the RESI estimates and confidence intervals. The interpretation of the RESI is the same as the coefficient table, but we note that in the ANOVA table, the RESI estimates are all nonnegative because they are estimated from F statistics. The estimates in the ANOVA table differ for two reasons: (1) Type II sum of squares first tests main effects without their interactions in the model; for example, the ``age'' RESI estimate is interpreted as the effect of age compared to a model that does not include age or the interaction term for age and region. (2) For variables that are tested on 1 degree of freedom, the ANOVA table estimates the absolute effect size, whereas the coefficient table uses the unbiased signed effect size by default (see \hyperref[sec:resi estimators]{RESI Estimators}). For example, with sex and BMI, we notice that the estimates are close in the ANOVA and coefficients tables, but not exactly equal in absolute value. This is due to using the default \code{unbiased = TRUE} argument, which uses the $t$ to S estimator~\eqref{eq:t2S} rather than the one based on the $F$ to S formula. 

An overall Wald test is also reported in the model. 

\begin{CodeChunk}
\begin{CodeInput}
R> resi_obj_lm$overall
\end{CodeInput}
\begin{CodeOutput}
Wald test

Model 1: charges ~ 1
Model 2: charges ~ region * age + sex + bmi
  Res.Df Df      F     Pr(>F)    RESI    2.5%   97.5%
1   1337                                             
2   1328  9 20.249 3.4435e-32 0.35954 0.32052 0.42138
\end{CodeOutput}
\end{CodeChunk}

By default, this compares the model to a reduced model that has only the intercept. The RESI estimate represents the overall absolute effect size of the model. In this model, this is estimated to be 0.3595 (CI: 0.3205, 0.4214). This is interpreted as a moderate to large effect.

\begin{figure}
    \centering
    \includegraphics{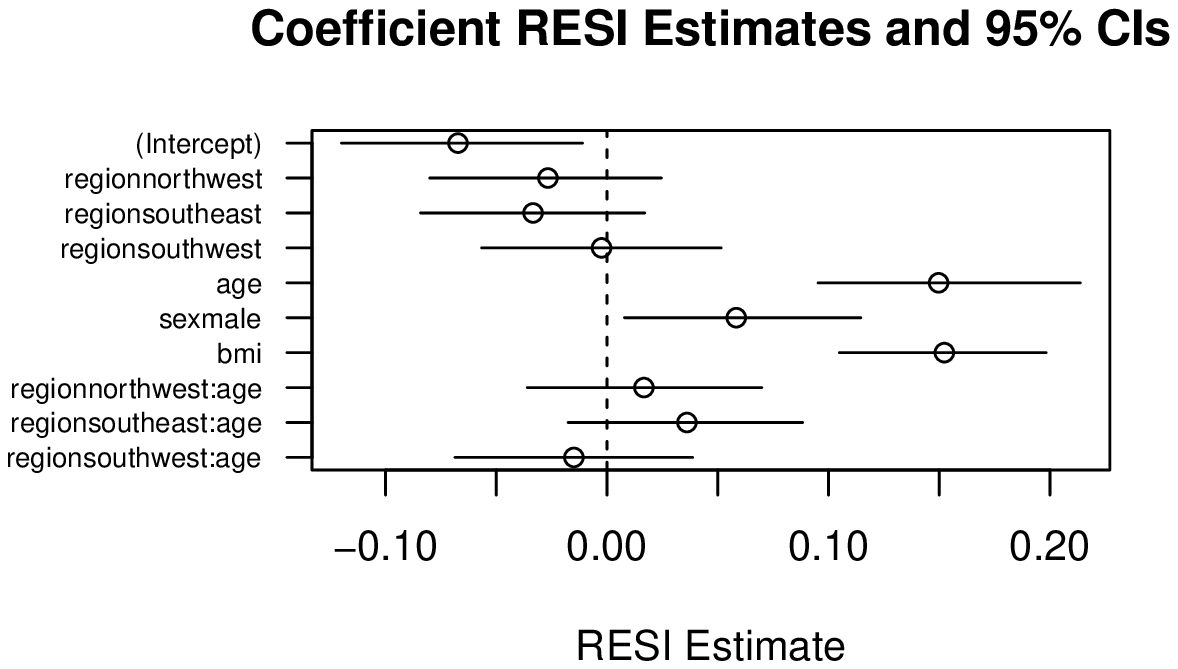}
    \includegraphics{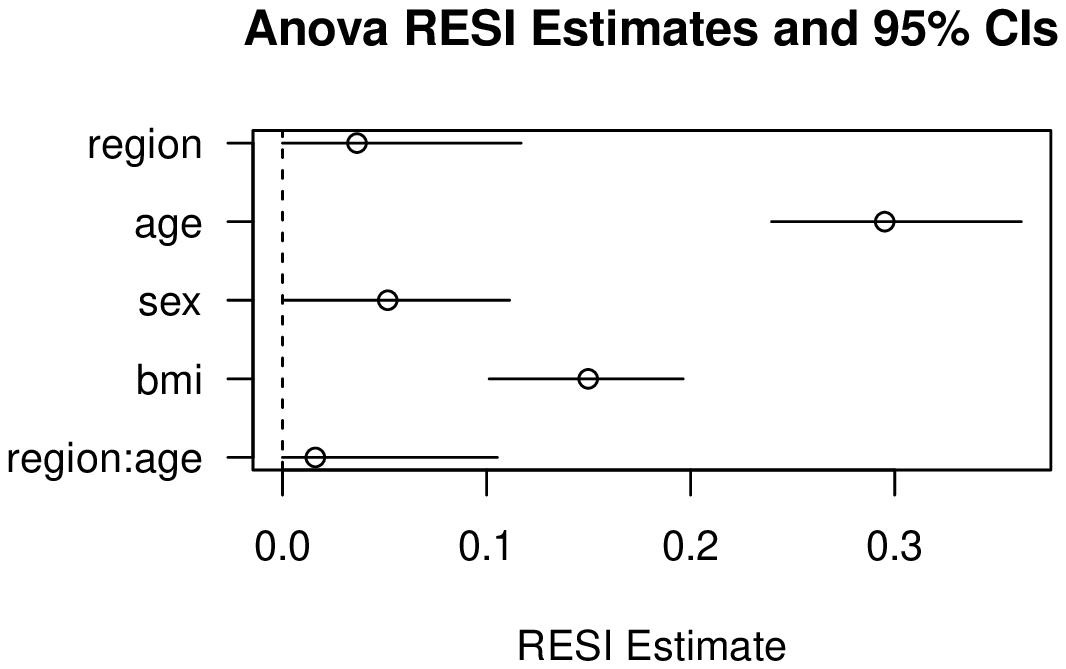}
    \caption{RESI estimates and confidence intervals from linear model coefficients and ANOVA tables.}
    \label{fig:ex1 plot}
\end{figure}

We can also visualize the results using the \code{plot()} function (Figure~\ref{fig:ex1 plot}). Running \code{plot()} on the `\code{resi}' object will plot the coefficient table. Because some of the variable names may be long, the user may want to adjust the margins before plotting. Another option is to use the \code{ycex.axis} argument in the \code{plot.resi()} method, which adjusts the text size of the of the y-axis relative to 1.
Alternatively, the user can extract the estimates and CIs from the tables and plot using their preferred visualization tool.

\begin{CodeChunk}
\begin{CodeInput}
R> par(mar = c(5, 7, 4, 2) + 0.1)
R> plot(resi_obj_lm, ycex.axis = 0.7)
R> plot(anova(resi_obj_lm))
\end{CodeInput}
\end{CodeChunk}

If we want to see a plot of the ANOVA table, we can run \code{plot()} either directly on the \code{anova} element of the `\code{resi}' object or on \code{anova()} or \code{car::Anova()} on the `\code{resi}' object. These plots help us quickly visualize the RESI estimates and relative effect sizes of the variables.

If we want to use different arguments for the covariance estimator function or the ANOVA function, we can specify these using the \code{vcov.args} and \code{Anova.args} arguments, respectively, in the \code{resi()} function. For example, we can use the \code{sandwich::vcovHC()} function with the HC0 estimator instead of the default HC3 estimator \citep{long_using_2000} and use Type III sum of squares instead of Type II as follows.

\begin{CodeChunk}
\begin{CodeInput}
R> set.seed(0826)
R> resi_obj_lm2 <- resi(mod_lm, vcov.args = list(type = "HC0"), 
R> Anova.args = list(type = 3))
R> resi_obj_lm2
\end{CodeInput}
\begin{CodeOutput}
Analysis of effect sizes based on RESI:
Confidence level =  0.05
Call:  lm(formula = charges ~ region * age + sex + bmi, data = insurance)

Coefficient Table 
                      Estimate Std. Error t value Pr(>|t|)    RESI    2.5%  
(Intercept)         -5359.4352  2155.7518 -2.4861   0.0130 -0.0679 -0.1210
regionnorthwest     -2339.4433  2372.3609 -0.9861   0.3243 -0.0269 -0.0808
regionsoutheast     -3230.8512  2618.6352 -1.2338   0.2175 -0.0337 -0.0849
regionsouthwest      -232.4839  2549.4981 -0.0912   0.9274 -0.0025 -0.0572
age                   220.3325    39.8159  5.5338   0.0000  0.1512  0.0964
sexmale              1328.0215   617.0486  2.1522   0.0316  0.0588  0.0080
bmi                   323.7725    57.5594  5.6250   0.0000  0.1537  0.1059
regionnorthwest:age    34.9040    56.6786  0.6158   0.5381  0.0168 -0.0363
regionsoutheast:age    83.6359    62.7223  1.3334   0.1826  0.0364 -0.0177
regionsouthwest:age   -33.6290    60.7822 -0.5533   0.5802 -0.0151 -0.0693
                         97.5%
(Intercept)            -0.0112
regionnorthwest         0.0248
regionsoutheast         0.0172
regionsouthwest         0.0520
age                     0.2157
sexmale                 0.1155
bmi                     0.2001
regionnorthwest:age     0.0706
regionsoutheast:age     0.0892
regionsouthwest:age     0.0390


Analysis of Deviance Table (Type III tests)

Response: charges
            Df       F Pr(>F)   RESI   2.5%  97.5%
(Intercept)  1  6.1807 0.0130 0.0622 0.0000 0.1179
region       3  0.7256 0.5367 0.0000 0.0000 0.0971
age          1 30.6228 0.0000 0.1487 0.0924 0.2139
sex          1  4.6320 0.0316 0.0521 0.0000 0.1122
bmi          1 31.6408 0.0000 0.1512 0.1023 0.1982
region:age   3  1.1388 0.3322 0.0175 0.0000 0.1065

Overall RESI comparing model to intercept-only model:

  Res.Df Df       F Pr(>F)   RESI   2.5%  97.5%
1   1328  9 20.6344      0 0.3631 0.3238 0.4255

Notes:
1. The RESI was calculated using a robust covariance estimator.
2. Confidence intervals (CIs) constructed using 1000 non-parametric
   bootstraps. 
\end{CodeOutput}
\end{CodeChunk}

Here, we print the full output of the `\code{resi}' object. In addition to elements previously discussed, notes on the type of covariance estimator (robust or naive) and type and number of bootstraps are found at the bottom. As expected, we can see that the results differ slightly from our first `\code{resi}' object output.

\subsection{RESI on nonlinear least squares} \label{illustration 2}

In this example, we use \code{resi()} on a nonlinear least squares model using \code{nls()}, demonstrating a helpful workaround to deal with model convergence issues in `\code{nls}' models when bootstrapping. For this analysis, we use the \code{niering} dataset in the \pkg{sars} package, available on CRAN \citep{matthews_sars_2019}. This dataset provides the area (in km$^2$) and number of plant species for 32 islands in the Kapingamarangi Atoll \citep{matthews_sars_2019}.

\begin{CodeChunk}
\begin{CodeInput}
R> data("niering", package = "sars")
R> head(niering)
\end{CodeInput}
\begin{CodeOutput}
        a  s
1 0.00012  5
2 0.00160  7
3 0.00240  8
4 0.00280 10
5 0.00360  9
6 0.00360 11
\end{CodeOutput}
\end{CodeChunk}

The species-to-area relationship is commonly modeled using a power curve, where $Species = cArea^z$ \citep{preston_canonical_1962}. We can fit this model using \code{nls()} to estimate the $c$ and $z$ parameters. It is a well known that `\code{nls}' models can be sensitive to the choice of starting values. For example, the following naive guesses for the starting values produce an error due to failed convergence.

\begin{CodeChunk}
\begin{CodeInput}
R> mod_nls <- nls(s ~ c*a^z, data = niering, start = list(c = 2, z = 0.5))
\end{CodeInput}
\begin{CodeOutput}
Error in nls(s ~ c * a^z, data = niering, start = list(c = 2, z = 0.5)) : 
singular gradient
\end{CodeOutput}
\end{CodeChunk}

If we use good starting values the model converges successfully.

\begin{CodeChunk}
\begin{CodeInput}
R> mod_nls <- nls(s ~ c*a^z, data = niering, start = list(c = 3, z = 0.25))
R> summary(mod_nls)
\end{CodeInput}
\begin{CodeOutput}
Formula: s ~ c * a^z

Parameters:
  Estimate Std. Error t value Pr(>|t|)    
c 89.30789   10.11148   8.832 7.59e-10 ***
z  0.40206    0.03677  10.935 5.49e-12 ***
---
Signif. codes:  0 '***' 0.001 '**' 0.01 '*' 0.05 '.' 0.1 ' ' 1

Residual standard error: 5.819 on 30 degrees of freedom

Number of iterations to convergence: 12 
Achieved convergence tolerance: 2.807e-06
\end{CodeOutput}
\end{CodeChunk}

With our `\code{nls}' model, we can run \code{resi()}, making sure to provide the data argument. For this example, we will demonstrate the Bayesian bootstrap option. 

\begin{CodeChunk}
\begin{CodeInput}
R> set.seed(0826)
R> resi_obj_nls <- resi(mod_nls, data = niering, boot.method = "bayes")
R> resi_obj_nls
\end{CodeInput}
\begin{CodeOutput}
Analysis of Effect sizes (ANOES) based on RESI:
Confidence level =  0.05
Call:  nls(formula = s ~ c * a^z, data = niering, start = list(c = 3, 
    z = 0.25), algorithm = "default", control = list(maxiter = 50, 
    tol = 1e-05, minFactor = 0.0009765625, printEval = FALSE, 
    warnOnly = FALSE, scaleOffset = 0, nDcentral = FALSE), trace = FALSE)

Coefficient Table 
  Estimate Std. Error t value Pr(>|t|)   RESI   2.5%  97.5%
c  89.3079    20.5866  4.3382    1e-04 0.7475 0.6206 1.5962
z   0.4021     0.0597  6.7325    0e+00 1.1601 0.9835 2.4238

Overall RESI comparing model to intercept-only model:

              chi2 df P   RESI  2.5%  97.5%
Wald Test 142.1953  2 0 2.0931 1.221 3.7403

Notes:
1. The RESI was calculated using a robust covariance estimator.
2. Credible intervals constructed using 1000 Bayesian bootstraps. 
3. The bootstrap was successful in 737 out of 1000 attempts.
\end{CodeOutput}
\end{CodeChunk}

The \code{resi()} function runs without error, and we obtain a coefficients table and an overall Wald test for the model with RESI estimates and 95\% credible intervals. Although the original model was able to be fit by \code{nls()} without issue, using this model for \code{resi()} does not have optimal performance. We can see from Note 3 that the bootstrap was only successful in 737 of the replicates.

% We can also obtain this information without printing the \code{resi} object with the \code{nfail} element.

% \begin{verbatim}
%     resi_obj_nls$nfail
% \end{verbatim}
% \verbatiminput{ex2 5.txt}

The unsuccessful replicates failed to converge when attempting to update the `\code{nls}' model with bootstrap data. We can improve the performance of \code{resi()} for this model by refitting the `\code{nls}' model with different start values before running \code{resi()}. We use the estimated coefficients from the original model as the new start values.

\begin{CodeChunk}
\begin{CodeInput}
R> mod_nls2 <- nls(s ~ c*a^z, data = niering, 
+               start = list(c = coef(mod_nlsn)[1], 
+               z = coef(mod_nlsn)[2]))
R> set.seed(0826)
R> resi(mod_nls2, data = niering, boot.method = "bayes")
\end{CodeInput}
\begin{CodeOutput}
Analysis of effect sizes based on RESI:
Confidence level =  0.05
Call:  nls(formula = s ~ c * a^z, data = niering, 
    start = list(c = coef(mod_nls)[1], z = coef(mod_nls)[2]), 
    algorithm = "default", control = list(maxiter = 50, 
    tol = 1e-05, minFactor = 0.0009765625, printEval = FALSE, 
    warnOnly = FALSE, scaleOffset = 0, nDcentral = FALSE), 
    trace = FALSE)

Coefficient Table 
    Estimate Std. Error t value Pr(>|t|)   RESI   2.5%  97.5%
c.c  89.3079    20.5866  4.3382    1e-04 0.7475 0.6178 1.9495
z.z   0.4021     0.0597  6.7325    0e+00 1.1601 0.9706 2.5139

Overall RESI comparing model to intercept-only model:

              chi2 df P   RESI   2.5%  97.5%
Wald Test 142.1953  2 0 2.0931 1.1893 3.6103

Notes:
1. The RESI was calculated using a robust covariance estimator.
2. Credible intervals constructed using 1000 Bayesian bootstraps. 
3. The bootstrap was successful in 1000 out of 1000 attempts.
\end{CodeOutput}
\end{CodeChunk}

We see that running \code{resi()} on this model gives us the same RESI estimates and similar credible intervals, but the performance of the bootstrap is much better. In this case all 1000 bootstrap replicates are successful, and we obtain credible intervals based on the desired number of bootstrap replicates. When using \code{resi()} on an `\code{nls}' model, consider using this strategy if the model fails to converge in many of the bootstrap samples.

\subsection{RESI on survival model} \label{illustration 3}

As a final example, we consider a parametric survival model using the \pkg{survival} package. Following an example in the survival package documentation, we fit a Weibull model using the \code{lung} dataset in the \pkg{survival} package \citep{therneau_package_2022}. The outcome is survival time (in days). The regressors are age, sex, and Karnofsky score.

It is important to note that for survival models (using \code{coxph()} or \code{survreg()}), the option to use a robust covariance is included in the model fitting function. The \code{resi()} function ignores the \code{vcovfunc} argument for these model types and assumes the user has specified the desired covariance method when fitting the model.

In this example we also demonstrate how the user can obtain confidence intervals for different levels of $\alpha$ both during and after running the \code{resi()} function. The \code{alpha} arguments allows the user to specify a vector of $\alpha$ levels, and the results corresponding to these levels will be output with the `\code{resi}' object. In the case that the user wants to produce different level confidence intervals after running the \code{resi()} function without rerunning the bootstrapping process the user can set \code{store.boot = TRUE}. This will store a data.frame in the `\code{resi}' object called \code{boot.results} that includes the all of the RESI estimates for each bootstrap replicate. Confidence intervals of a specific $\alpha$ level can then be obtained manually or using the \code{summary()} or \code{anova()}/\code{car::Anova()} functions. 

For this example we will use the \code{unbiased = FALSE} option to demonstrate the alternate z to S estimator described in equation~\eqref{eq:z2S_alt}. We also specify a reduced model to compute a RESI for a subset of the model parameters, rather than using an intercept-only model. Our reduced model uses Karnofsky score as the only predictor and we use 1500 bootstrap replicates to construct CIs.

\begin{CodeChunk}
\begin{CodeInput}
R> library("survival")
R> set.seed(0826)
R> mod_surv <- survreg(Surv(time, status) ~ age + sex + ph.karno,
+                       data = survival::lung, dist="weibull", 
+                       robust = TRUE)
R> mod_surv_reduced <- survreg(Surv(time, status) ~ ph.karno,
+                     data = survival::lung, dist="weibull", 
+                     robust = TRUE)
R> resi_obj_surv <- resi(mod_surv, mod_surv_reduced, data = survival::lung, 
+                   unbiased = FALSE, store.boot = TRUE,
+                   alpha = c(0.05, 0.1), nboot = 1500)
R> resi_obj_surv
\end{CodeInput}
\begin{CodeOutput}
Analysis of effect sizes based on RESI:
Confidence level =  0.05 0.1
Full Model:survreg(formula = Surv(time, status) ~ age + sex + ph.karno, 
    data = survival::lung, dist = "weibull", robust = TRUE)
Reduced Model:survreg(formula = Surv(time, status) ~ ph.karno, 
    data = survival::lung, dist = "weibull", robust = TRUE)


Coefficient Table 
            Estimate Std. Error z value Pr(>|z|)    RESI    2.5%      5%    
(Intercept)   5.3263     0.6854  7.7711   0.0000  0.5104  0.3304  0.3602 
age          -0.0089     0.0073 -1.2174   0.2235 -0.0460 -0.2016 -0.1799 
sex           0.3702     0.1225  3.0216   0.0025  0.1888  0.0283  0.0689 
ph.karno      0.0093     0.0058  1.5873   0.1124  0.0816  0.0000  0.0000 
Log(scale)   -0.2808     0.0674 -4.1643   0.0000 -0.2677 -0.4564 -0.4236
                 95%   97.5%
(Intercept)   0.6765  0.7167
age           0.0000  0.0000
sex           0.3048  0.3200
ph.karno      0.2762  0.3146
Log(scale)   -0.1668 -0.1381

Analysis of Deviance Table (Type II tests)

Response: Surv(time, status)
         Df  Chisq Pr(>Chisq)   RESI   2.5%     5%    95%  97.5%
age       1 1.4820     0.2235 0.0460 0.0000 0.0000 0.1799 0.2016
sex       1 9.1299     0.0025 0.1888 0.0283 0.0689 0.3048 0.3200
ph.karno  1 2.5196     0.1124 0.0816 0.0000 0.0000 0.2762 0.3146

Overall RESI comparing full model to reduced model:

  Res.Df Df  Chisq Pr(>Chisq) RESI   2.5%     5%    95% 97.5%
1    222  2 10.232      0.006 0.19 0.0315 0.0899 0.3233 0.343

Notes:
1. The RESI was calculated using a robust covariance estimator.
2. Confidence intervals (CIs) constructed using 1500 non-parametric 
bootstraps. 
\end{CodeOutput}
\end{CodeChunk}

The printed output reflects the modifications we made to the \code{resi()} arguments. The reduced model formula is displayed, which is relevant only for the overall RESI estimate. For comparison, we can look at the \code{overall} element of running \code{resi()} with an intercept-only reduced model.

\begin{CodeChunk}
\begin{CodeInput}
R> set.seed(0826)
R> resi(mod_surv, data = survival::lung, 
+    unbiased = FALSE, alpha = c(0.05, 0.1), nboot = 1500)$overall
\end{CodeInput}
\begin{CodeOutput}
Wald test

Model 1: Surv(time, status) ~ 1
Model 2: Surv(time, status) ~ age + sex + ph.karno
  Res.Df Df  Chisq Pr(>Chisq)    RESI   2.5%      5%     95%   97.5%
1    225                                                            
2    222  3 11.544  0.0091     0.1936 0.0678  0.1034  0.3774  0.4091
\end{CodeOutput}
\end{CodeChunk}

The overall RESI estimate is slightly higher when comparing to an intercept-only model than the model that adjusts for Karnofsky score.

The coefficient table and ANOVA table are computed only for the full model. Because we chose the unbiased option, the RESI estimates are equal in absolute value for the coefficient and ANOVA tables. The RESI estimates for age and Karnofsky (-0.0460 (95\% CI: -0.2016, 0) and 0.0816 (95\% CI: 0, 0.3146) respectively) are interpreted as small effects, while the RESI estimate for sex (0.1888 (95\% CI: 0.0283, 0.3200)) is interpreted as a small to moderate effect. We see from the output that there are now four columns for the RESI confidence intervals - a lower and upper bound for each of the $\alpha$ levels specified. If we now want to obtain an interval with a different confidence level, we can run \code{summary()} and \code{anova()} using the \code{alpha} argument, and specify a vector of values.

\begin{CodeChunk}
\begin{CodeInput}
R> summary(resi_obj_surv, alpha = c(0.001, 0.01))
\end{CodeInput}
\begin{CodeOutput}
Analysis of effect sizes based on RESI:
Confidence level =  0.001 0.01
Call:  survreg(formula = Surv(time, status) ~ age + sex + ph.karno, 
    data = survival::lung, dist = "weibull", robust = TRUE)

Coefficient Table 
            Estimate Std. Error z value Pr(>|z|)    RESI   0.05%    0.5%
(Intercept)   5.3263     0.6854  7.7711   0.0000  0.5104  0.2427  0.2800
age          -0.0089     0.0073 -1.2174   0.2235 -0.0460 -0.2952 -0.2412
sex           0.3702     0.1225  3.0216   0.0025  0.1888  0.0000  0.0000
ph.karno      0.0093     0.0058  1.5873   0.1124  0.0816 -0.0437  0.0000
Log(scale)   -0.2808     0.0674 -4.1643   0.0000 -0.2677 -0.6118 -0.5037
               99.5%  99.95%
(Intercept)  0.7725  0.9481
age          0.0560  0.1055
sex          0.3661  0.4645
ph.karno     0.3896  0.4496
Log(scale)  -0.1003 -0.0728
\end{CodeOutput}
\end{CodeChunk}

\begin{CodeChunk}
\begin{CodeInput}
R> anova(resi_obj_surv, alpha = c(0.001, 0.01))
\end{CodeInput}
\begin{CodeOutput}
          Df  Chisq Pr(>Chisq)     RESI 0.05% 0.5%   99.5%  99.95%
age        1 1.4820   0.223460 0.045979     0    0 0.24120 0.29524
sex        1 9.1299   0.002515 0.188832     0    0 0.36612 0.46448
ph.karno   1 2.5196   0.112441 0.081638     0    0 0.38961 0.44956
\end{CodeOutput}
\end{CodeChunk}

Note that if we try to specify different $\alpha$ levels with these functions on a `\code{resi}' object that did not use the \code{store.boot = TRUE} option, an error will occur with a message informing the user that this option was not used.
A larger number of bootstrap samples are necessary to obtain adequate precision for smaller \code{alpha} levels.

\section{Conclusion}

The \pkg{RESI} \proglang{R} package aims to provide estimates and confidence intervals for the recently introduced index in a way that intuitively complements common data analysis workflow in R. Similarly to running \code{summary()} after fitting a model, a user can simply run \code{resi()} on many models and obtain several useful model summaries that include both $p$~values and RESI estimates with confidence intervals. There are dedicated methods for several common model types currently, with more in process. Methods for both cross-sectional and longitudinal models are available, with longitudinal methods providing both a longitudinal and a per-measurement cross-sectional RESI estimate. For models that are not currently implemented, users can manually provide the relevant information to functions within \pkg{RESI} to obtain estimates directly. The package also makes it easy to visualize RESI estimates and convert to and from other effect size indices. The RESI is a widely applicable effect size index with several advantages, including the ability to accommodate nuisance parameters and incorporate robust covariance estimates. With increasing emphasis being placed on reporting of effect sizes in research, the \pkg{RESI} package is a user-friendly tool to easily report effect sizes and confidence intervals in publications.

\section{Computational details}

All examples were coded using \proglang{R} version 4.2.2 and \pkg{RESI} version 1.1.0 \citep{r_core_team_r_2022, jones_resi_2023}. The versions of relevant packages for the examples include \pkg{sandwich} 3.0-2 \citep{zeileis_object-oriented_2006}, \pkg{sars} 1.3.6 \citep{matthews_sars_2019}, and \pkg{survival} 3.5-3 \citep{therneau_package_2022}.

%\section{Data Availability} \label{data}

%There are three datasets used in this article. The \code{insurance} dataset is available in the \text{RESI} package and adapted from the following repository:
%Kaggle. US Health Insurance Dataset. %\url{https://www.kaggle.com/datasets/teertha/ushealthinsurancedataset}.
%This project contains the ``insurance.csv" data file that  is available under the terms of the %\href{https://creativecommons.org/publicdomain/zero/1.0/}{CC0: Public Domain dedication}.

%The \code{niering} dataset is available in the \code{sars} package \citep{matthews_sars_2019}. The data can be found directly at \url{https://github.com/txm676/sars/blob/master/data/niering.rda}. The data was originally provided in \citep{niering_terrestrial_1963}.

%The \code{lung} dataset is provided in the \code{survival} package \citep{therneau_package_2022}. The data were originally provided in \citep{loprinzi_prospective_1994}.

%\section{Software Availability}

%\begin{itemize}
    %\item \textit{Software available from:} \url{https://cran.r-project.org/package=RESI}
    %\item \textit{Source code available from:} \url{https://github.com/statimagcoll/RESI}
    %\item \textit{Archived source code at time of publication:} DOI and citation for project in Zenodo (please select the appropriate DOI for the version which underlies your article).
    %\item \textit{License:} \href{https://cran.r-project.org/web/licenses/GPL-3}{GPL-3}
%\end{itemize}

%\section{Competing Interests}

%No competing interests were disclosed.

\section{Acknowledgements}

This research is funded by R01MH123563.

\bibliography{./MyLibrary}

\begin{thebibliography}{50}
\newcommand{\enquote}[1]{``#1''}
\providecommand{\natexlab}[1]{#1}
\providecommand{\url}[1]{\texttt{#1}}
\providecommand{\urlprefix}{URL }
\expandafter\ifx\csname urlstyle\endcsname\relax
  \providecommand{\doi}[1]{doi:\discretionary{}{}{}#1}\else
  \providecommand{\doi}{doi:\discretionary{}{}{}\begingroup
  \urlstyle{rm}\Url}\fi
\providecommand{\eprint}[2][]{\url{#2}}

\bibitem[{Agresti(2002)}]{agresti_categorical_2002}
Agresti A (2002).
\newblock \emph{Categorical {Data} {Analysis}}.
\newblock Wiley {Series} in {Probability} and {Statistics}. John Wiley \& Sons,
  Inc., Hoboken, NJ, USA.
\newblock ISBN 978-0-471-36093-3 978-0-471-24968-9.
\newblock \doi{10.1002/0471249688}.
\newblock \urlprefix\url{http://doi.wiley.com/10.1002/0471249688}.

\bibitem[{Althouse \emph{et~al.}(2021)Althouse, Below, Claggett, Cox, de~Lemos,
  Deo, Duval, Hachamovitch, Kaul, Keith, Secemsky, Teixeira-Pinto, Roger, and
  null}]{althouse_recommendations_2021}
Althouse AD, Below JE, Claggett BL, Cox NJ, de~Lemos JA, Deo RC, Duval S,
  Hachamovitch R, Kaul S, Keith SW, Secemsky E, Teixeira-Pinto A, Roger VL,
  null n (2021).
\newblock \enquote{Recommendations for {Statistical} {Reporting} in
  {Cardiovascular} {Medicine}: {A} {Special} {Report} {From} the {American}
  {Heart} {Association}.}
\newblock \emph{Circulation}, \textbf{144}(4), e70--e91.
\newblock \doi{10.1161/CIRCULATIONAHA.121.055393}.
\newblock Publisher: American Heart Association,
  \urlprefix\url{https://www.ahajournals.org/doi/10.1161/CIRCULATIONAHA.121.055393}.

\bibitem[{Amaral and Line(2021)}]{amaral_current_2021}
Amaral EdOS, Line SRP (2021).
\newblock \enquote{Current use of effect size or confidence interval analyses
  in clinical and biomedical research.}
\newblock \emph{Scientometrics}, \textbf{126}(11), 9133--9145.
\newblock ISSN 0138-9130.
\newblock \doi{10.1007/s11192-021-04150-3}.

\bibitem[{{American Psychological
  Association}(2001)}]{american_psychological_association_publication_2001}
{American Psychological Association} (2001).
\newblock \emph{Publication manual of the {American} {Psychological}
  {Association}.}
\newblock 5th ed. edition. American Psychological Association, Washington, DC.
\newblock ISBN 978-1-55798-791-4.

\bibitem[{{American Psychological
  Association}(2010)}]{american_psychological_association_publication_2010}
{American Psychological Association} (2010).
\newblock \emph{Publication manual of the {American} {Psychological}
  {Association}.}
\newblock 6th ed. edition. American Psychological Association, Washington, DC.
\newblock ISBN 978-1-4338-0561-5.

\bibitem[{Anderson(2020)}]{anderson_esvis_2020}
Anderson D (2020).
\newblock \enquote{esvis: {Visualization} and {Estimation} of {Effect}
  {Sizes}.}
\newblock \urlprefix\url{https://CRAN.R-project.org/package=esvis}.

\bibitem[{Association(1994)}]{american_psychological_association_publication_1994}
Association AP (1994).
\newblock \emph{Publication manual of the {American} {Psychological}
  {Association}, 4th ed.}
\newblock Publication manual of the {American} {Psychological} {Association},
  4th ed. American Psychological Association, Washington, DC, US.
\newblock ISBN 1-55798-243-0 (Hardcover); 1-55798-241-4 (Paperback).
\newblock Pages: xxxii, 368.

\bibitem[{Bates \emph{et~al.}(2015)Bates, Mächler, Bolker, and
  Walker}]{bates_fitting_2015}
Bates D, Mächler M, Bolker B, Walker S (2015).
\newblock \enquote{Fitting {Linear} {Mixed}-{Effects} {Models} {Using} lme4.}
\newblock \emph{Journal of Statistical Software}, \textbf{67}(1), 1--48.
\newblock ISSN 1548-7660.
\newblock \doi{10.18637/jss.v067.i01}.
\newblock
  \urlprefix\url{https://www.jstatsoft.org/index.php/jss/article/view/v067i01}.

\bibitem[{Ben-Shachar \emph{et~al.}(2020)Ben-Shachar, Lüdecke, and
  Makowski}]{ben-shachar_effectsize_2020}
Ben-Shachar M, Lüdecke D, Makowski D (2020).
\newblock \enquote{effectsize: {Estimation} of {Effect} {Size} {Indices} and
  {Standardized} {Parameters}.}
\newblock \emph{Journal of Open Source Software}, \textbf{5}(56), 2815.
\newblock ISSN 2475-9066.
\newblock \doi{10.21105/joss.02815}.
\newblock \urlprefix\url{https://joss.theoj.org/papers/10.21105/joss.02815}.

\bibitem[{Betensky(2019)}]{betensky_p-value_2019}
Betensky RA (2019).
\newblock \enquote{The p-{Value} {Requires} {Context}, {Not} a {Threshold}.}
\newblock \emph{The American Statistician}, \textbf{73}(sup1), 115--117.
\newblock ISSN 0003-1305.
\newblock \doi{10.1080/00031305.2018.1529624}.
\newblock Publisher: Taylor \& Francis \_eprint:
  https://doi.org/10.1080/00031305.2018.1529624,
  \urlprefix\url{https://doi.org/10.1080/00031305.2018.1529624}.

\bibitem[{Boos and Stefanski(2013)}]{boos_essential_2013}
Boos DD, Stefanski LA (2013).
\newblock \emph{Essential {Statistical} {Inference}: {Theory} and {Methods}}.
\newblock Springer {Texts} in {Statistics}. Springer-Verlag, New York.
\newblock ISBN 978-1-4614-4817-4.
\newblock \urlprefix\url{//www.springer.com/us/book/9781461448174}.

\bibitem[{Buchanan \emph{et~al.}(2019)Buchanan, Gillenwaters, Scofield, and
  Valentine}]{buchanan_mote_2019}
Buchanan EM, Gillenwaters A, Scofield JE, Valentine KD (2019).
\newblock \emph{{MOTE}: {Measure} of the {Effect}: {Package} to assist in
  effect size calculations and their confidence intervals}.
\newblock \urlprefix\url{http://github.com/doomlab/MOTE}.

\bibitem[{Carey(2022)}]{carey_gee_2022}
Carey VJ (2022).
\newblock \emph{gee: {Generalized} {Estimation} {Equation} {Solver}}.
\newblock \urlprefix\url{https://CRAN.R-project.org/package=gee}.

\bibitem[{Cohen(1988)}]{cohen_statistical_1988}
Cohen J (1988).
\newblock \emph{Statistical power analysis for the behavioral sciences}.
\newblock Erlbaum Associates, Hillsdale, NJ.

\bibitem[{Fox and Weisberg(2019)}]{fox_r_2019}
Fox J, Weisberg S (2019).
\newblock \emph{An {R} {Companion} to {Applied} {Regression}}.
\newblock Third edition. Sage, Thousand Oaks CA.
\newblock
  \urlprefix\url{https://socialsciences.mcmaster.ca/jfox/Books/Companion/}.

\bibitem[{Fritz \emph{et~al.}(2012)Fritz, Morris, and
  Richler}]{fritz_effect_2012}
Fritz CO, Morris PE, Richler JJ (2012).
\newblock \enquote{Effect size estimates: {Current} use, calculations, and
  interpretation.}
\newblock \emph{Journal of Experimental Psychology: General}, \textbf{141}(1),
  2--18.
\newblock ISSN 1939-2222(Electronic),0096-3445(Print).
\newblock \doi{10.1037/a0024338}.

\bibitem[{Halekoh \emph{et~al.}(2006)Halekoh, Højsgaard, and
  Yan}]{halekoh_r_2006}
Halekoh U, Højsgaard S, Yan J (2006).
\newblock \enquote{The {R} {Package} geepack for {Generalized} {Estimating}
  {Equations}.}
\newblock \emph{Journal of Statistical Software}, \textbf{15/2}, 1--11.

\bibitem[{Hedges and Olkin(1985)}]{hedges_statistical_1985}
Hedges LV, Olkin I (1985).
\newblock \emph{Statistical {Methods} for {Meta}-{Analysis}}.
\newblock Elsevier, London, UK.
\newblock ISBN 978-0-08-057065-5.
\newblock \doi{10.1016/C2009-0-03396-0}.
\newblock
  \urlprefix\url{https://linkinghub.elsevier.com/retrieve/pii/C20090033960}.

\bibitem[{Jackman(2020)}]{jackman_pscl_2020}
Jackman S (2020).
\newblock \emph{pscl: {Classes} and {Methods} for {R} {Developed} in the
  {Political} {Science} {Computational} {Laboratory}}.
\newblock United States Studies Centre, University of Sydney, Sydney, New South
  Wales, Australia.
\newblock \urlprefix\url{https://github.com/atahk/pscl/}.

\bibitem[{Jones \emph{et~al.}(2023)Jones, Kang, and Vandekar}]{jones_resi_2023}
Jones M, Kang K, Vandekar S (2023).
\newblock \emph{{RESI}: {Robust} {Effect} {Size} {Index} ({RESI})
  {Estimation}}.
\newblock \urlprefix\url{https://CRAN.R-project.org/package=RESI}.

\bibitem[{Kang \emph{et~al.}(2023)Kang, Jones, Armstrong, Avery, McHugo,
  Heckers, and Vandekar}]{kang_accurate_2023}
Kang K, Jones MT, Armstrong K, Avery S, McHugo M, Heckers S, Vandekar S (2023).
\newblock \enquote{Accurate {Confidence} and {Bayesian} {Interval} {Estimation}
  for {Non}-centrality {Parameters} and {Effect} {Size} {Indices}.}
\newblock \emph{Psychometrika}.
\newblock ISSN 1860-0980.
\newblock \doi{10.1007/s11336-022-09899-x}.

\bibitem[{Kelley(2022)}]{kelley_mbess_2022}
Kelley K (2022).
\newblock \enquote{{MBESS}: {The} {MBESS} {R} {Package}.}
\newblock \urlprefix\url{https://CRAN.R-project.org/package=MBESS}.

\bibitem[{{Lesnoff} \emph{et~al.}(2012){Lesnoff}, {M.}, {Lancelot}, and
  {R.}}]{lesnoff_aod_2012}
{Lesnoff}, {M}, {Lancelot}, {R} (2012).
\newblock \emph{aod: {Analysis} of {Overdispersed} {Data}}.
\newblock \urlprefix\url{https://cran.r-project.org/package=aod}.

\bibitem[{Long and Ervin(2000)}]{long_using_2000}
Long JS, Ervin LH (2000).
\newblock \enquote{Using heteroscedasticity consistent standard errors in the
  linear regression model.}
\newblock \emph{The American Statistician}, \textbf{54}(3), 217--224.

\bibitem[{MacKinnon and
  White(1985)}]{mackinnon_heteroskedasticity-consistent_1985}
MacKinnon JG, White H (1985).
\newblock \enquote{Some heteroskedasticity-consistent covariance matrix
  estimators with improved finite sample properties.}
\newblock \emph{Journal of econometrics}, \textbf{29}(3), 305--325.

\bibitem[{Mangiafico(2023)}]{mangiafico_rcompanion_2023}
Mangiafico S (2023).
\newblock \emph{rcompanion: {Functions} to {Support} {Extension} {Education}
  {Program} {Evaluation}}.
\newblock \urlprefix\url{https://CRAN.R-project.org/package=rcompanion}.

\bibitem[{Mantel(1963)}]{mantel_chi-square_1963}
Mantel N (1963).
\newblock \enquote{Chi-{Square} {Tests} with {One} {Degree} of {Freedom};
  {Extensions} of the {Mantel}- {Haenszel} {Procedure}.}
\newblock \emph{Journal of the American Statistical Association},
  \textbf{58}(303), 690--700.
\newblock ISSN 0162-1459.
\newblock \doi{10.2307/2282717}.
\newblock Publisher: [American Statistical Association, Taylor \& Francis,
  Ltd.], \urlprefix\url{https://www.jstor.org/stable/2282717}.

\bibitem[{Matloff and Yancey(2022)}]{matloff_regtools_2022}
Matloff N, Yancey R (2022).
\newblock \emph{regtools: {Regression} and {Classification} {Tools}}.
\newblock \urlprefix\url{https://CRAN.R-project.org/package=regtools}.

\bibitem[{Matthews \emph{et~al.}(2019)Matthews, Triantis, Whittaker, and
  Guilhaumon}]{matthews_sars_2019}
Matthews TJ, Triantis K, Whittaker RJ, Guilhaumon F (2019).
\newblock \enquote{sars: an {R} package for fitting, evaluating and comparing
  species–area relationship models.}
\newblock \emph{Ecography}, \textbf{42}, 1446--1455.

\bibitem[{Papachristodoulou and Prajna(2005)}]{papachristodoulou_tutorial_2005}
Papachristodoulou A, Prajna S (2005).
\newblock \enquote{A tutorial on sum of squares techniques for systems
  analysis.}
\newblock In \emph{Proceedings of the 2005, {American} {Control} {Conference},
  2005.}, pp. 2686--2700. IEEE.

\bibitem[{Pinheiro \emph{et~al.}(2021)Pinheiro, Bates, DebRoy, Sarkar, and {R
  Core Team}}]{pinheiro_nlme_2021}
Pinheiro J, Bates D, DebRoy S, Sarkar D, {R Core Team} (2021).
\newblock \emph{nlme: {Linear} and {Nonlinear} {Mixed} {Effects} {Models}}.
\newblock \urlprefix\url{https://CRAN.R-project.org/package=nlme}.

\bibitem[{Preston(1962)}]{preston_canonical_1962}
Preston FW (1962).
\newblock \enquote{The {Canonical} {Distribution} of {Commonness} and {Rarity}:
  {Part} {I}.}
\newblock \emph{Ecology}, \textbf{43}(2), 185.
\newblock ISSN 00129658.
\newblock \doi{10.2307/1931976}.
\newblock \urlprefix\url{http://www.jstor.org/stable/1931976?origin=crossref}.

\bibitem[{Pustejovsky(2022)}]{pustejovsky_clubsandwich_2022}
Pustejovsky J (2022).
\newblock \emph{{clubSandwich}: {Cluster}-{Robust} ({Sandwich}) {Variance}
  {Estimators} with {Small}-{Sample} {Corrections}}.
\newblock \urlprefix\url{https://CRAN.R-project.org/package=clubSandwich}.

\bibitem[{{R Core Team}(2022)}]{r_core_team_r_2022}
{R Core Team} (2022).
\newblock \emph{R: {A} {Language} and {Environment} for {Statistical}
  {Computing}}.
\newblock R Foundation for Statistical Computing, Vienna, Austria.
\newblock \urlprefix\url{https://www.R-project.org/}.

\bibitem[{Rosenthal(1994)}]{rosenthal_parametric_1994}
Rosenthal R (1994).
\newblock \enquote{Parametric measures of effect size.}
\newblock \emph{The handbook of research synthesis}, \textbf{621}, 231--244.

\bibitem[{Rubin(1981)}]{rubin_bayesian_1981}
Rubin DB (1981).
\newblock \enquote{The {Bayesian} {Bootstrap}.}
\newblock \emph{The Annals of Statistics}, \textbf{9}(1), 130--134.
\newblock ISSN 0090-5364, 2168-8966.
\newblock \doi{10.1214/aos/1176345338}.
\newblock
  \urlprefix\url{https://projecteuclid.org/journals/annals-of-statistics/volume-9/issue-1/The-Bayesian-Bootstrap/10.1214/aos/1176345338.full}.

\bibitem[{Serdar \emph{et~al.}(2021)Serdar, Cihan, Yücel, and
  Serdar}]{serdar_sample_2021}
Serdar CC, Cihan M, Yücel D, Serdar MA (2021).
\newblock \enquote{Sample size, power and effect size revisited: simplified and
  practical approaches in pre-clinical, clinical and laboratory studies.}
\newblock \emph{Biochemia Medica}, \textbf{31}(1), 010502.
\newblock ISSN 1330-0962.
\newblock \doi{10.11613/BM.2021.010502}.
\newblock
  \urlprefix\url{https://www.ncbi.nlm.nih.gov/pmc/articles/PMC7745163/}.

\bibitem[{Sullivan and Feinn(2012)}]{sullivan_using_2012}
Sullivan GM, Feinn R (2012).
\newblock \enquote{Using {Effect} {Size}—or {Why} the {P} {Value} {Is} {Not}
  {Enough}.}
\newblock \emph{Journal of Graduate Medical Education}, \textbf{4}(3),
  279--282.
\newblock ISSN 1949-8349.
\newblock \doi{10.4300/JGME-D-12-00156.1}.
\newblock \urlprefix\url{https://doi.org/10.4300/JGME-D-12-00156.1}.

\bibitem[{Therneau(2022)}]{therneau_package_2022}
Therneau T (2022).
\newblock \enquote{A package for survival analysis in {R}.}

\bibitem[{Torchiano(2020)}]{torchiano_effsize_2020}
Torchiano M (2020).
\newblock \emph{effsize: {Efficient} {Effect} {Size} {Computation}}.
\newblock \doi{10.5281/zenodo.1480624}.
\newblock \urlprefix\url{https://CRAN.R-project.org/package=effsize}.

\bibitem[{Vandekar \emph{et~al.}(2020)Vandekar, Tao, and
  Blume}]{vandekar_robust_2020}
Vandekar S, Tao R, Blume J (2020).
\newblock \enquote{A {Robust} {Effect} {Size} {Index}.}
\newblock \emph{Psychometrika}, \textbf{85}(1), 232.

\bibitem[{Vandekar and Stephens(2021)}]{vandekar_improving_2021}
Vandekar SN, Stephens J (2021).
\newblock \enquote{Improving the replicability of neuroimaging findings by
  thresholding effect sizes instead of p-values.}
\newblock \emph{Human Brain Mapping}, \textbf{42}(8), 2393--2398.
\newblock ISSN 1097-0193.
\newblock \doi{https://doi.org/10.1002/hbm.25374}.
\newblock
  \urlprefix\url{https://onlinelibrary.wiley.com/doi/abs/10.1002/hbm.25374}.

\bibitem[{Wasserstein and Lazar(2016)}]{wasserstein_asas_2016}
Wasserstein RL, Lazar NA (2016).
\newblock \enquote{The {ASA}’s statement on p-values: context, process, and
  purpose.}
\newblock \emph{The American Statistician}, \textbf{70}(2), 129--133.

\bibitem[{White(1980)}]{white_heteroskedasticity-consistent_1980}
White H (1980).
\newblock \enquote{A heteroskedasticity-consistent covariance matrix estimator
  and a direct test for heteroskedasticity.}
\newblock \emph{Econometrica: Journal of the Econometric Society}, pp.
  817--838.

\bibitem[{Wickham \emph{et~al.}(2022)Wickham, Hester, Chang, and
  Bryan}]{wickham_devtools_2022}
Wickham H, Hester J, Chang W, Bryan J (2022).
\newblock \emph{devtools: {Tools} to {Make} {Developing} {R} {Packages}
  {Easier}}.
\newblock \urlprefix\url{https://CRAN.R-project.org/package=devtools}.

\bibitem[{Wilkinson(1999)}]{wilkinson_statistical_1999}
Wilkinson L (1999).
\newblock \enquote{Statistical methods in psychology journals: {Guidelines} and
  explanations.}
\newblock \emph{American Psychologist}, \textbf{54}, 594--604.
\newblock ISSN 1935-990X(Electronic),0003-066X(Print).
\newblock \doi{10.1037/0003-066X.54.8.594}.
\newblock Place: US Publisher: American Psychological Association.

\bibitem[{Zeileis(2006)}]{zeileis_object-oriented_2006}
Zeileis A (2006).
\newblock \enquote{Object-oriented {Computation} of {Sandwich} {Estimators}.}
\newblock \emph{Journal of Statistical Software}, \textbf{16}, 1--16.
\newblock ISSN 1548-7660.
\newblock \doi{10.18637/jss.v016.i09}.
\newblock \urlprefix\url{https://doi.org/10.18637/jss.v016.i09}.

\bibitem[{Zeileis and Hothorn(2002)}]{zeileis_diagnostic_2002}
Zeileis A, Hothorn T (2002).
\newblock \enquote{Diagnostic {Checking} in {Regression} {Relationships}.}
\newblock \emph{R News}, \textbf{2}(3), 7--10.
\newblock \urlprefix\url{https://CRAN.R-project.org/doc/Rnews/}.

\bibitem[{Zeileis \emph{et~al.}(2020)Zeileis, Köll, and
  Graham}]{zeileis_various_2020}
Zeileis A, Köll S, Graham N (2020).
\newblock \enquote{Various {Versatile} {Variances}: {An} {Object}-{Oriented}
  {Implementation} of {Clustered} {Covariances} in {R}.}
\newblock \emph{Journal of Statistical Software}, \textbf{95}(1), 1--36.
\newblock \doi{10.18637/jss.v095.i01}.

\bibitem[{Zhang and Schoeps(1997)}]{zhang_robust_1997}
Zhang Z, Schoeps N (1997).
\newblock \enquote{On robust estimation of effect size under semiparametric
  models.}
\newblock \emph{Psychometrika}, \textbf{62}(2), 201--214.
\newblock ISSN 1860-0980.
\newblock \doi{10.1007/BF02295275}.
\newblock \urlprefix\url{https://doi.org/10.1007/BF02295275}.

\end{thebibliography}

\end{document}